\documentclass[prc,notitlepage, twocolumn,
showpacs,floatfix,nofootinbib,preprintnumbers,superscriptaddress,amsmath,amssymb]{revtex4-1}
\usepackage{graphicx}
\usepackage{dcolumn}
\usepackage{bm}
\usepackage[usenames,dvipsnames]{color}
\usepackage[normalem]{ulem} 

\newcommand{\mkrm}[1]{}                                      
\newcommand{\HFBTHO}{\textsc{hfbtho}}

\renewcommand{\vec}[1]{\mbox{\boldmath $#1$}}

\begin{document}

\title{Multipole modes in deformed  nuclei within the finite amplitude method}

\author{M. Kortelainen}%
\affiliation{%
University of Jyvaskyla, Department of Physics, P.O. Box 35, FI-40014 University of Jyvaskyla, Finland
}
\affiliation{%
Helsinki Institute of Physics, P.O. Box 64, FI-00014 University of Helsinki, Finland
}
\author{N. Hinohara}%
\affiliation{%
Center for Computational Sciences, University of Tsukuba, Tsukuba 305-8577, Japan
}
\affiliation{%
NSCL/FRIB Laboratory, Michigan State University, East Lansing, Michigan 48824, USA
}
\author{W. Nazarewicz}%
\affiliation{%
Department of Physics and Astronomy and NSCL/FRIB Laboratory, Michigan State University, East Lansing, Michigan 48824, USA
}
\affiliation{%
Institute of Theoretical Physics, Faculty of Physics, University of Warsaw, PL-02-093 Warsaw, Poland
}


\begin{abstract}
\begin{description}
\item[Background]
To access selected excited states of nuclei, within the framework of nuclear density functional theory,
the quasiparticle random phase approximation (QRPA) is commonly used.
\item[Purpose]
We present a computationally efficient, fully self-consistent framework to compute the QRPA transition strength function of
an arbitrary multipole operator in  axially-deformed superfluid nuclei.
\item[Methods]
The method is based on the finite amplitude method (FAM) QRPA, allowing fast iterative solution of  QRPA equations. A numerical implementation of the FAM-QRPA solver module has been carried out for deformed nuclei.
\item[Results]
The practical feasibility of the deformed FAM module has been  demonstrated.
In particular, we calculate the quadrupole and octupole  strength  in a heavy deformed nucleus $^{240}$Pu, without
any truncations in the quasiparticle space. To demonstrate the capability to calculate individual QRPA modes, we also compute low-lying negative-parity collective states in $^{154}$Sm.
\item[Conclusions] 
The new FAM implementation enables calculations of the QRPA  strength function throughout the nuclear landscape. This will facilitate global surveys of multipole modes and beta decays, and will open new avenues for constraining the nuclear energy density functional. 
\end{description}

\end{abstract}

\maketitle


{\it Introduction} --
The response of the atomic nucleus to an external perturbation provides valuable
information about the underlying nuclear structure and characteristics of the
nuclear force~\cite{Bohr-MottelsonV2,(Lip89),(rin00),(Paa07)}. In addition to nuclear physics aspects, 
electromagnetic excitations and transition rates have a profound impact on r-process and stellar 
nucleosynthesis~\cite{(Arn07)}. Theoretically, a  microscopic 
description of a system with hundreds of strongly interacting fermions is a challenging 
task. Because exact {\it ab-initio} methods are still  computationally out of  reach for open-shell, heavy  systems, self-consistent mean-field models
rooted in nuclear density functional theory (DFT)  are usually employed 
when it comes to  complex deformed nuclei \cite{(rin00),(Ben03)}.
The main ingredient of the nuclear DFT is the energy density functional (EDF). Current
EDF models have demonstrated the ability
to provide a fairly accurate description of nuclear ground state properties across 
the nuclear chart, despite local deficiencies \cite{(Ben03),(Erl12),(Bog13etal),(Kor14)}.

To access the excited states of nucleus in the framework of nuclear DFT, one of the most 
straightforward and commonly used method is the linear response theory within  
random-phase-approximation (RPA) or quasiparticle random-phase-approximation (QRPA).
Traditionally, the nuclear QRPA problem has been formulated in a matrix form (MQRPA).
Due to large dimension of QRPA matrices, especially when spherical symmetry is  broken, 
fully self-consistent deformed MQRPA calculations have become possible only recently \cite{(Art08),(Ter10),(Los10),(Yos10),(Per08),(Per11),(Yos11),(Mar11),(Ter11),(Mus13)}.
The large computational cost of deformed MQRPA implies that
various truncations of quasi-particle space must be introduced.
Such cut-offs, however, break the self-consistency 
between the underlying Hartree-Fock-Bogoliubov (HFB) solution and  QRPA, and may cause an appearance of spurious states.

In order to circumvent various practical deficiencies  of MQRPA, a finite amplitude method (FAM) was introduced
as a way to compute multipole strength function. 
With FAM, the QRPA problem is solved iteratively, avoiding costly computation of the
MQRPA matrix elements and a subsequent diagonalization. 
It was first implemented for a computation of the RPA strength function \cite{(Nak07)}, and then applied
to a spherically symmetric QRPA~\cite{(Avo11)}.
In the work of Ref.~\cite{(Sto11)} the FAM-QRPA was extended to the axially symmetric case within the Skyrme-HFB framework in harmonic oscillator basis.
The feasibility of FAM in the framework of relativistic mean field models was studied for the spherical~\cite{(Lia13)} and  axially-symmetric \cite{(Nik13)} cases. Recently, FAM was also used together with  an axially symmetric coordinate-space HFB solver~\cite{(Pei14)}.

The FAM turned out to be a versatile theoretical tool with a broad  range of applications in addition to    strength function evaluations. For instance,  it was demonstrated that it can be used to compute the MQRPA matrix~\cite{(Avo13)};  individual QRPA modes \cite{(Hin13)};
sum-rules \cite{(Hin15)}; and $\beta$ decay rates \cite{(Mus14)}. An alternative to FAM to solve the QRPA problem iteratively is the iterative Arnoldi diagonalization scheme, which solves the QRPA equations in a reduced Krylov space~\cite{(Toi10)}. This method was  also applied to 
superfluid systems and discrete QRPA states~\cite{(Ves12),(Car12)}.

The objective of this work is to extend the FAM to the deformed case, allowing evaluation  QRPA modes for  operators of arbitrary multipolarity $LK$. This is an extension of our earlier work~\cite{(Sto11)} that was limited to  $K=0$.


{\it Theoretical framework} --
Our formulation of the FAM-QRPA directly follows that of Ref.~\cite{(Sto11)} where details can be found. The FAM equations can be written as:
\begin{subequations}
\begin{eqnarray}
\left(E_{\mu}+E_{\nu}-\omega\right)X_{\mu\nu}(\omega) + \delta H^{20}_{\mu\nu}(\omega) & = & -F^{20}_{\mu\nu}, \\
\left(E_{\mu}+E_{\nu}+\omega\right)Y_{\mu\nu}(\omega) + \delta H^{02}_{\mu\nu}(\omega) & = & -F^{02}_{\mu\nu},
\end{eqnarray}
\end{subequations}
where $F^{20}$ and $F^{02}$ are constructed from the external multipole  field $f$ that perturbs the system,  and
$X_{\mu\nu}(\omega)$ and $Y_{\mu\nu}(\omega)$ are the FAM-QRPA amplitudes at a given excitation energy 
$\omega$. Furthermore, $\delta H^{20}(\omega)$ and $\delta H^{02}(\omega)$ define the response of the nucleus to the 
external field~\cite{(Sto11)}.

In the original formulation of the FAM, the induced fields were calculated by taking a numerical derivative
with respect of a small expansion parameter $\eta$:
$\delta h(\omega)     =(h[\rho_\eta,\kappa_\eta,\bar{\kappa}_\eta]-h[\rho,\kappa,\kappa^*])/\eta$,  
$\delta \Delta (\omega)     =  (\Delta[\rho_\eta,\kappa_\eta]-\Delta[\rho,\kappa])/\eta$, and $\overline{\delta  \Delta}(\omega)  =  (\Delta[\bar{\rho}_\eta,\bar{\kappa}_\eta]-\Delta[\rho,\kappa])/\eta,$
where $\rho$ and $\kappa$ are the HFB particle density and pair density (pairing tensor), respectively, and 
$\rho_\eta$, $\bar{\rho}_\eta$, $\kappa_\eta$, and $\bar{\kappa}_\eta$ are the corresponding FAM densities  that depend
on  $\eta$. 
In the $K\ne 0$ case considered here, however, the coordinate-space fields 
$h, \Delta$, and $\overline{\Delta}$
must be linearized explicitly
in order not to mix densities with different values of the magnetic quantum number $K$. Such a linearization is possible since the oscillating
part of the density, proportional to $\eta$, is assumed to be small compared to the static HFB density.
Due to this explicit linearization, the expansion parameter $\eta$ is no longer needed and the induced densities are:
\begin{subequations}
\begin{eqnarray}
\rho_{\rm f}          & = &  + UXV^{T}               + V^{*}Y^{T}U^{\dag}, \\
\bar{\rho}_{\rm f}    & = &  + V^{*}X^{\dag}U^{\dag} + UY^{*}V^{T}, \\
\kappa_{\rm f}        & = &  - UX^{T}U^{T}           - V^{*}YV^{\dag}, \\
\bar{\kappa}_{\rm f}  & = &  - V^{*}X^{*}V^{\dag}    - UY^{\dag}U^{T}  \, ,
\label{eq:rhoetano2}
\end{eqnarray}
\end{subequations}
where $U$ and $V$ are the usual HFB matrices, and the subscript f indicates oscillating densities induced by the external field atop of the static HFB density. The linearized fields are:
$\delta h(\omega)  = h[\rho_{\rm f},\kappa_{\rm f},\bar{\kappa}_{\rm f}]$,
$\delta \Delta (\omega)        = \Delta[\rho_{\rm f},\kappa_{\rm f}]$, and
$\overline{\delta  \Delta}(\omega) =  \Delta[\bar{\rho}_{\rm f},\bar{\kappa}_{\rm f}].$
In practice, for Skyrme-like EDFs, the explicit linearization is required for the density-dependent fields.

In implementation of the new FAM module, we have utilized the simplex-$y$ ($\hat{S}_{y}$) 
symmetry \cite{(Dob00)}.  Consequently, the basis states used are eigenstates of
$\hat{S}_{y}$ operator corresponding to eigenvalues of  $+i$ and
$-i$; they can be written as combinations of $|+\Omega\rangle$ and
$|-\Omega\rangle$ states, where $\Omega$ is the  projection of the single-particle angular momentum along  the $z$-axis \cite{(Goo74)}.
With a proper selection of the operator $f$ for the external field, basis states with opposite simplex eigenvalues are 
not connected by the induced density matrix $\rho_{\rm f}$. 
In a $K\ne 0$ case, the density matrix has a block structure, dictated by the operator $f$,
corresponding to the selection rule $\Delta \Omega =  K$.

In terms of FAM-QRPA amplitudes, the multipole strength can be expresses as:
\begin{equation}
\frac{dB(\omega;F)}{d\omega} = -\frac{1}{\pi} {\rm Im\,Tr} \left[ f(UXV^{T}+V^{*}Y^{T}U^{\dag}) \right].
\end{equation}
To guarantee that the FAM-QRPA solution has finite  strength, a small imaginary 
component is introduced to the excitation energy $\omega$ as $\omega\,\to\,\omega+i\gamma$ \cite{(Nak07)}.
Actually, the position of $\omega$ in the complex plane does not need to be limited to this particular choice: by choosing a  suitable  integration contour in the complex-$\omega$ plane, discrete QRPA states or sum rules can be obtained \cite{(Hin13),(Hin15)}.

The electric isoscalar (IS) and isovector (IV) multipole operators are \cite{(Lip89)}:
\begin{equation}
f^{\rm IS}_{LK} =  e_{\rm IS} \sum_{i=1}^{A} {f}_{LK}(\vec{r}_i),~
f^{\rm IV}_{LK}  =  \sum_{i=1}^{A} e_{{\rm IV},\tau_i}\tau_i{f}_{LK}(\vec{r}_i),
\end{equation}
where $\tau_i=\pm1$ for neutrons/protons, $f_{LK}(\vec{r})=r^L {Y}_{LK}(\hat{\vec{r}})$,
and $e_{\rm IS}$ and $e_{{\rm IV},\tau_i}$ are isoscalar and isovector effective charges, respectively. As simplex-$y$ is considered to be a self-consistent symmetry,  
 one can replace
\begin{equation}
f_{LK} \rightarrow f^{+}_{LK} =\left( f_{LK} + f_{L,-K}\right)/\sqrt{2-\delta_{K0}} \label{eq:fplus}
\end{equation}
and assume  $K\ge 0$ in the following. Indeed, for an even-even axial nucleus,
operators $f_{LK}$ and $f_{L,-K}$ produce identical strength functions.

Our FAM-QRPA implementation is based on the DFT code {\HFBTHO} \cite{(Sto13)}, which
solves the HFB equations in axially symmetric (transformed) harmonic oscillator basis by assuming 
time-reversal symmetry. 
The iterative Broyden method of Ref.~\cite{(Bar08)} is used to speed up the convergence of the FAM-QRPA iterations.
For the direct Coulomb part, we use the same method as in the version v200d of {\HFBTHO} \cite{(Sto13)}, generalized to the $K\ne 0$ case. We benchmarked the new FAM code against the old FAM module of Ref.~\cite{(Sto11)} in the case of monopole and quadrupole modes with  $K=0$, and obtained perfect agreement.
For the negative-parity electric operators, the used coordinate mesh also included the half-volume corresponding to  negative-$z$ values.

We would like to stress that, unlike in the standard deformed MQRPA, we do not impose any kind of
truncation on the quasiparticle FAM-QRPA space. The only cut-off (besides the size of the harmonic oscillator basis) is
the employed pairing window used for the calculation of induced densities, in order to keep self-consistency with respect to the underlying HFB calculation.

The calculation of the FAM strength function can be trivially parallelized by distributing parts of the strength function
over multiple CPU cores. To this end, we have implemented a parallel MPI calculation scheme. In practice, a computation of a typical strength function with 20 oscillator shells, and without the reflection symmetry assumed, on a multicore Intel Sandy Bridge 2.6\,GHz processor system,
takes about 1000 CPU hours.


{\it Results} --
In our illustrative examples, we have used two Skyrme EDF parameterizations, SkM*~\cite{(Bar82)} and SLy4~\cite{(Cha95)}.
Both parameterizations have been found to be stable to linear response in infinite nuclear matter
\cite{(Pas12)}. 


In a spherical nucleus, the strength function for 
a given multipole $L$ does not depend on $K$ quantum number. This offers a stringent test of our  numerical 
implementation of the FAM module.
To this end, we computed the  isovector quadrupole strength for $^{20}$O, 
with SkM* Skyrme EDF, in a space of $N_{\rm sh}=15$ oscillator shells, by using mixed pairing interaction with
strength of $V_0=-280\,{\rm MeV\,fm^{3}}$ and a quasiparticle cut-off of  50\,MeV.
The setup of this calculation was the same as in the  MQRPA calculation of Ref.~\cite{(Los10)} to facilitate comparison.
We confirmed that  the transition strengths of all $K$-modes coincide, and the results
agree very well with those of Ref.~\cite{(Los10)}.
The relative differences between various $K$-modes in our calculations were typically at the level of  $\mathcal{O}(10^{-5})$, or smaller.

\begin{figure}[!htb]
\includegraphics[width=\linewidth]{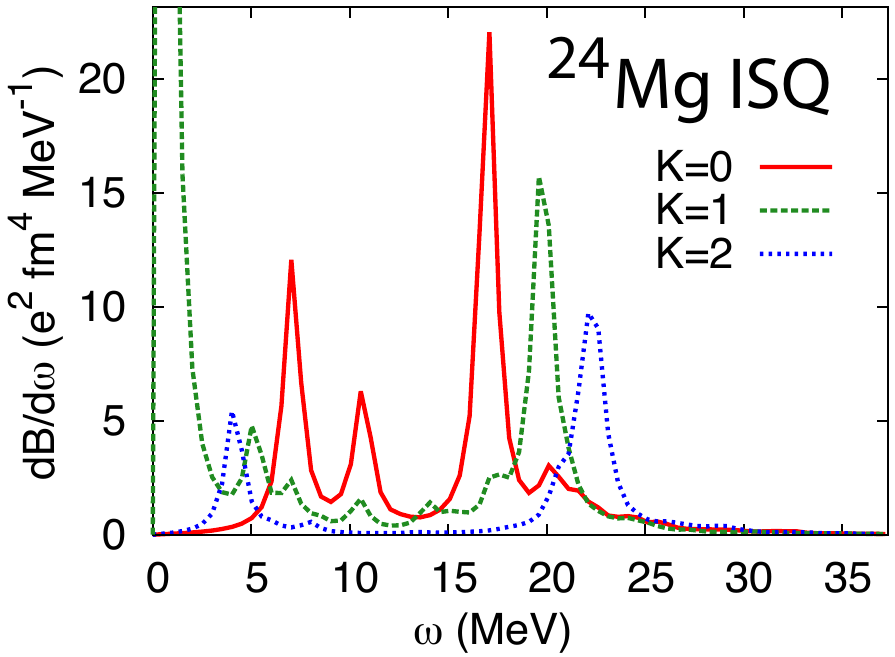}
\caption{(Color online)  Isoscalar quadrupole  strength in the prolate deformed configuration
of $^{24}$Mg  calculated with SkM* EDF and $N_{\rm sh}=15$.}
\label{fig:24Mgisq}
\end{figure}
Figure~\ref{fig:24Mgisq}  shows the calculated isoscalar quadrupole transition strength in $^{24}$Mg.
The calculation was done by using the same setup as in the case of $^{20}$O. Here, we consider the deformed configuration of  $^{24}$Mg with
quadrupole deformation  $\beta=0.39$. In this configuration, static pairing  vanishes for both protons and neutrons.
Due to the deformation,  strength functions of different $K$-modes differ.
By comparing our results with those of Ref.~\cite{(Los10)}, we again find  excellent agreement,
except  for the spurious reorientation Nambu-Goldstone $K=1$ mode  that shows up just above $\omega=0$.  For more discussion of spurious modes in FAM-QRPA we refer the reader to the recent
paper \cite{(Hin15b)}.


\begin{figure}[!htb]
\center
\includegraphics[width=\linewidth]{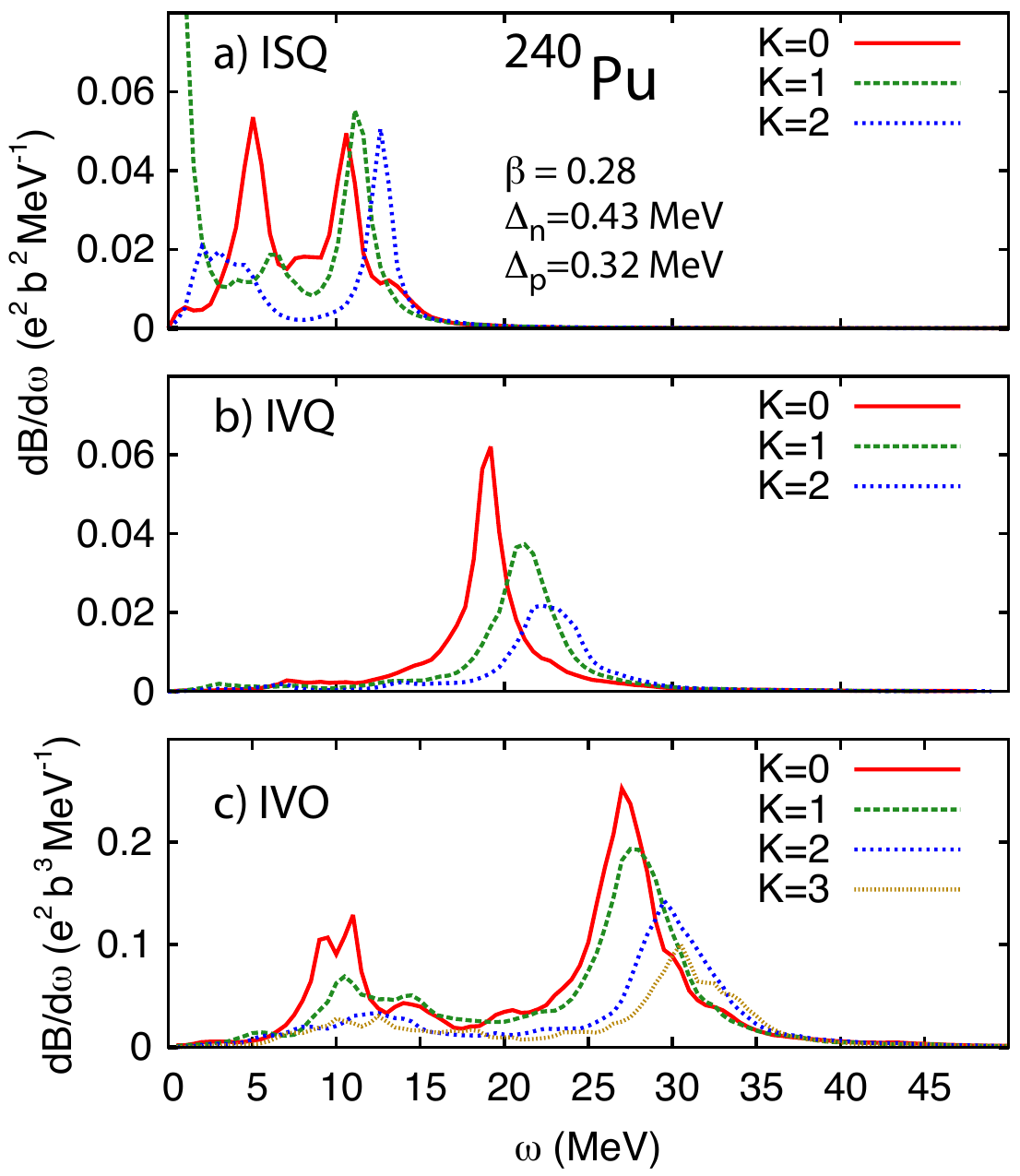}
\caption{(Color online) Isoscalar (a) and isovector (b) quadrupole  strengths, and isovector octupole  strength (c) in 
  $^{240}$Pu calculated with SLy4  EDF and $N_{\rm sh}=20$.}
\label{fig:240Pu}
\end{figure}
To demonstrate the performance of the new FAM module for  deformed heavy nuclei, we calculated the quadrupole and octupole transition strengths
in $^{240}$Pu. The results obtained with 20 oscillator shells are presented in Fig.~\ref{fig:240Pu},
which shows a typical pattern dominated by the presence of giant quadrupole (GQR) and giant octupole (GOR) resonances. 
In this case we used SLy4   EDF together with a mixed pairing force with a strength of $V_{0}=-283.45\,{\rm MeV\,fm^{3}}$. 
The resulting HFB state had deformation $\beta=0.28$, and pairing gaps $\Delta_{\rm n}=0.43\,{\rm MeV}$ and 
$\Delta_{\rm p} = 0.32\,{\rm MeV}$. 

Our calculations predict  the $K$-splitting of the multipole strength due to deformation. 
For the IS-GQR, the splitting follows the pattern predicted by phenomenological models
\cite{(Kis75),(Lip81),(Nis85)}, i.e., for the prolate deformations the ISGQR energy increases with $K$.
A similar hierarchy is predicted for IV-GQR and IV-GOR.  
The mean  GQR  energies shown in Figs.~\ref{fig:240Pu}(a) and (b)  are consistent with the values predicted in the recent time-dependent DFT calculations of Ref.~\cite{(Sca14)} and the MQRPA study of Ref.~\cite{(Per11)}. The latter work also contains predictions for the octupole response in the neighboring nucleus $^{238}$U. Similar as in Fig.~~\ref{fig:240Pu}(c), they predict a strong fragmentation of low-energy and high-energy octupole strength. The mean energy of the high-energy IVGOR predicted in our work, around 28\,MeV, agrees well with the early predictions of Ref.~\cite{(Mal76)}.
Once again, for the isoscalar quadrupole mode with $K=1$, we find a spurious state related to the rotational Nambu-Goldstone 
mode. In addition, we have also tested that, by using  a stretched harmonic oscillator basis, the new FAM module can be employed to compute the multipole  strength in the fission isomer of $^{240}$Pu.

\begin{figure}[!htb]
\center
\includegraphics[width=\linewidth]{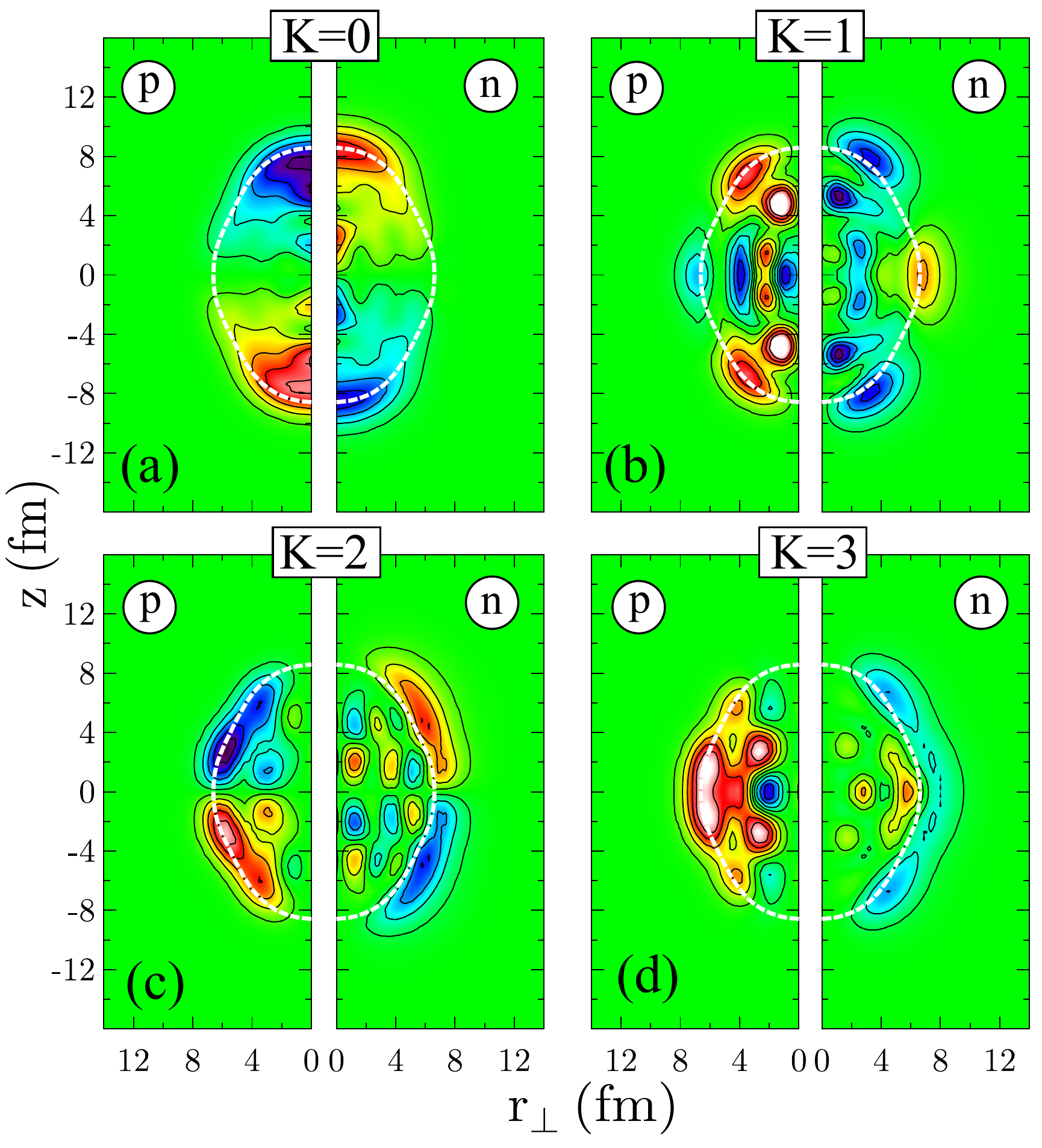}
\caption{(Color online) The imaginary part of the induced IVO transition density $\rho_{\rm f}$ 
at the excitation energy $\omega=11$\,MeV 
for protons and neutrons in $^{240}$Pu. All $K$-modes have been normalized in the same way:
Red color indicates the maximum (positive) value for each mode and  blue color indicates the minimum (negative) value.
The white dashed line indicates the contour of $\rho_{\rm n}+\rho_{\rm p}= 0.08\,{\rm fm}^{-3}$ obtained from the HFB calculation.}
\label{fig:240densi}
\end{figure}
To shed light on the spatial structure of induced transition density, we show in Fig.~\ref{fig:240densi}
the induced proton and neutron IVO transition densities in $^{240}$Pu, for all the $K$-modes,
at $\omega=11$\,MeV. Owing  to the isovector character of  the mode, protons and neutrons exhibit
out-of-phase oscillations. 
Furthermore, the spatial transition densities show a clear octupole pattern. The transition densities  cover a significant portion 
of the nuclear volume; this reflects the collective character of the mode.

Finally, we demonstrate the capability of the new FAM module to compute the discrete QRPA modes. 
The samarium and neodymium isotopes around $A=150$ are known to exhibit low-energy octupole modes. 
We have chosen an octupole-stable isotope $^{154}$Sm and calculated the low-lying octupole  vibrational states, using the same computational setup as for $^{240}$Pu.
The ground-state quadrupole deformation predicted in HFB  was $\beta=0.32$, and the pairing gaps were $\Delta_{\rm n}=0.30$\,MeV and $\Delta_{\rm p}=0.53$\,MeV.
The calculation was carried out by using the contour integration technique of Ref.~\cite{(Hin13)} and by applying an external
isovector octupole field to extract the individual states.
To confirm our results, we  repeated the  calculations by using the isovector dipole ($K=0$ and 1) and isoscalar octupole ($K=2$ and 3) fields. 
Table~\ref{table:154Sm} displays the isovector octupole transition strengths and corresponding proton $B$(E3) values. The $K=0$ and 1 
excited states carry the octupole strength that is larger than 1 W.u., indicating their collective nature.

Experimentally, two negative parity rotational bands with the band heads of $J^\pi=1^-$, 921.3\,keV and $J^\pi=1^-$, 1475.8\,keV have 
been identified in $^{154}$Sm. Those bands have been associated with  $K^\pi=0^-$ and $1^-$ octupole vibrations, respectively.
Although our calculation underestimates the experimental excitation energies of these   states, the 
$B$(E3) value of the $K^\pi=0^-$ state agrees well with the experimental value $B$(E3;$0_1^+\to3_1^-)=10(2)$ W.u. \cite{(spe89)}.
The excitation energies of the lowest $K^\pi=0^-$ and $1^-$ excited states in $^{154}$Sm are also presented in Ref.~\cite{(Yos11)}, and their values obtained with MQRPA with SkM* EDF are higher than ours.
The translational spurious modes appear at $\omega=0.11$\,MeV ($K^\pi=0^-$) and $\pm$0.17$i$\,MeV ($K^\pi=1^-$), and since the lowest 
$K^\pi=0^-$ collective state is  close to the spurious mode, some contamination due to the spurious components is expected. We are in the process of implementing  the prescription proposed in  Ref.~\cite{(Nak07)} to remove the spurious components from FAM-QRPA modes. 
\begin{table}
\caption{Lowest octupole QRPA modes in $^{154}$Sm predicted in our deformed FAM calculations. Shown are: 
the energy $\omega_1$;  the IVO transition strength $|\langle 0|f^{IV,+}_{L=3,K}|1\rangle|^2$; and the  corresponding $B$(E3) value. The transition probabilities were computed through the QRPA amplitudes (referred to as FAM-C in \cite{(Hin13)}). \label{table:154Sm}}
  \begin{ruledtabular}
    \begin{tabular}{cccc}
      $K$ & $\omega_1$ & $|\langle 0|f^{IV,+}_{L=3,K}|1\rangle|^2$ & $B$(E3)  \\
        & (MeV) &  ($e^2$fm$^6$ MeV$^{-1})$ & (W.u.) \\ \hline
      0 & 0.4168 & 6.684 & 8.70 \\
      1 & 0.9014 & 69.74 & 2.01 \\ 
      2 & 2.5973 & 1.916 & 0.24 \\ 
      3 & 1.3155 & 0.01809 & 0.0004
    \end{tabular}
\end{ruledtabular}
\end{table}


{\it Conclusions} -- 
In this work we have introduced  the FAM-QRPA method  suitable for calculation of an arbitrary
 multipole strength function in axially deformed superfluid nuclei. The method allows a fast
calculation of the  strength function without any additional truncations in the quasiparticle space. The method has been benchmarked in spherical and deformed nuclei by comparing with earlier MQRPA calculations~\cite{(Los10)}.
To demonstrate the applicability of the method to heavy deformed nuclei, we calculated quadrupole and octupole
strength functions in $^{240}$Pu. We also showed that the deformed FAM module can be used
to compute discrete QRPA modes.

Since the majority of nuclei are predicted to be axially deformed in their ground states, the proposed FAM-QRPA method is a  tool of the choice to 
study the linear multipole response across the nuclear landscape. Large-scale surveys with the deformed FAM-QRPA approach can be carried out very efficiently as the method is amendable to  parallel computing. Another useful application is in the area of EDF optimization, where new experimental information on multipole strength in deformed nuclei can be used to better constrain the isovector sector of the effective interaction.

\begin{acknowledgments}
We are grateful to Jacek Dobaczewski for helpful comments.
This material is
based upon work supported by Academy of Finland under the Centre of Excellence Programme 2012--2017 
(Nuclear and Accelerator Based Physics Programme at JYFL) and FIDIPRO programme;
and by the U.S.\ Department of Energy, Office of
Science, Office of Nuclear Physics under  award numbers 
DE-SC0013365 (Michigan State University) and
DE-SC0008511 (NUCLEI SciDAC-3).
We acknowledge
the CSC-IT Center for Science Ltd., Finland,
High Performance Computing Center, Institute for Cyber-Enables Research, Michigan State University, USA, and
COMA (PACS-IX) System at the Center for Computational Sciences, University of Tsukuba, Japan,
for the allocation of computational resources.
\end{acknowledgments}

\bibliographystyle{apsrev4-1}

\end{document}